%% file: journal.tex
\documentclass[conference]{IEEEtran}
\IEEEoverridecommandlockouts

\usepackage{graphicx}
\usepackage{url}
\usepackage{cite}
\usepackage{multicol}
\usepackage{multirow}
\usepackage{listings}
\usepackage{makecell}
\usepackage{xcolor}
\usepackage{color}
\usepackage{amssymb}
\usepackage{pifont}
\usepackage[linesnumbered,ruled,vlined]{algorithm2e}
\usepackage{booktabs}
\usepackage{pifont}

\SetKwInput{KwInput}{Input}                
\SetKwInput{KwOutput}{Output}              

\newenvironment{packed_itemize}{
    \begin{list}{\labelitemi}{\leftmargin=1.1em}
    \setlength{\itemsep}{4pt}
    \setlength{\parskip}{0pt}
    \setlength{\parsep}{0pt}
    \setlength{\headsep}{0pt}
    \setlength{\topskip}{0pt}
    \setlength{\topmargin}{0pt}
    \setlength{\topsep}{0pt}
    \setlength{\partopsep}{0pt}
}{\end{list}}

\newcommand{\sys}{{RoBin}\xspace}

\newcommand{\etal}{\mbox{\emph{et al.\ }}}
\newcommand{\ie}{\mbox{\emph{i.e.,\ }}}
\newcommand{\eg}{\mbox{\emph{e.g.,\ }}}
\newcommand{\cmark}{\ding{51}}
\newcommand{\xmark}{\ding{55}}

\begin{document}
%
\title{\sys : Facilitating the Reproduction of Configuration-Related Vulnerability}


\author{
\IEEEauthorblockN{Ligeng Chen\IEEEauthorrefmark{2}, Jian Guo\IEEEauthorrefmark{2}, Zhongling He\IEEEauthorrefmark{2}, Dongliang Mu\IEEEauthorrefmark{4}$^*$
\thanks{* Corresponding author.} and Bing Mao\IEEEauthorrefmark{2}} 
  
\IEEEauthorblockA{\IEEEauthorrefmark{2}National Key Laboratory for Novel Software Technology, Nanjing University, Nanjing, China \\ 
\{chenlg, zhe\}@smail.nju.edu.cn, jaguo2014@outlook.com, maobing@nju.edu.cn}

\IEEEauthorblockA{\IEEEauthorrefmark{4}School of Cyber Science and Engineering, Huazhong University of Science and Technology, Wuhan, China\\ dzm91@hust.edu.cn}}

\maketitle
\thispagestyle{empty}

\begin{abstract}

Vulnerability reproduction paves a way in debugging software failures, which need intensive manual efforts. However, some key factors (\eg software configuration, trigger method) are often missing, so we can not directly reproduce the failure without extra attempts. Even worse, highly customized configuration options of programs create a barrier for reproducing the vulnerabilities that only appear under some specific combinations of configurations.

In this paper, we address the problem mentioned above --- reproducing the configuration-related vulnerability. We try to solve it by proposing a binary similarity-based method to infer the specific building configurations via the binary from crash report. 
The main challenges are as follows:
precise compilation option inference, program configuration inference, and source-code-to-binary matching. To achieve the goal, we implement \sys, a binary similarity-based building configuration inference tool. To demonstrate the effectiveness, we test \sys on 21 vulnerable cases upon 4 well-known open-source programs. It shows a strong ability in pinpointing the building configurations causing the vulnerability. The result can help developers reproduce and diagnose the vulnerability, and finally, patch the programs. 
\end{abstract}

\begin{IEEEkeywords}
Binary Code Similarity, Building Configurations, Reproduction, Vulnerability.
\end{IEEEkeywords}

\IEEEpeerreviewmaketitle

\input{intro}
\input{background}

\input{analysis}
\input{definition}
\input{overview}

\input{design}
\input{eval}
\input{related}
\input{conclusion}





\bibliographystyle{ieeetr}
\bibliography{ref}





\end{document}

%% file: intro.tex
\section{Introduction} 

Despite the best efforts of software developers, software systems inevitably contain defects that may be leveraged as vulnerabilities. In the real world, these vulnerabilities could lead to many notorious cyberattacks, such as Stuxnet, WannaCry Ransomware~\cite{wannacry}, HeartBleed~\cite{heartbleed}. Since modern software systems are becoming more complex and release cycles are getting shorter, testing teams are unable to identify all the possible vulnerabilities before a software release. Thus software systems are typically released with many underlying vulnerabilities and end-users may experience software failures. Considering the performance overhead, when software crashes, only a crash report with the necessary information will be sent to the developers. The developers first need to reproduce the failure so can they diagnose and patch the vulnerability.




\textbf{Existing Works.} For software failures, the reproduction of underlying vulnerabilities is a key stage in the vulnerability diagnosis. Without reproduction, software developers are unable to diagnose the root causes and eventually verify whether vulnerabilities are fixed or not. There are several existing works that attempt to reproduce the underlying vulnerabilities and enables software developers to check the input and program state that lead to software failures. Mu et al.~\cite{Reproduction} show that vulnerability reproduction is difficult in the real world and there are many missing key factors in the vulnerability reports, such as building configurations, Proof-of-Concept (PoC), trigger method, etc. REPT~\cite{weidong2018osdi} leverages online trace recording from Intel Processor Tracing (Intel PT) and offline binary analysis to facilitate reverse debugging of software failures. Hercules~\cite{hercules} takes advantage of symbolic execution to generate one input which reproduces the give software crash. BugRedux~\cite{bugredux} conducted one empirical analysis, to understand the trade-offs between different execution data recording and the effectiveness of approaches.

\textbf{Limitations.} 
However, some vulnerabilities only occur when the specific code segment controlled by specific program configurations is compiled into the binary, which we call \textbf{\textit{configuration-related vulnerabilities}}. Only can the developers reproduce the failure when they get the binary containing the vulnerable code segment.
Nevertheless, though they can reproduce the failure, software developers and security analysts still need to instrument source code with sanitizers (\eg Address Sanitizer~\cite{asan}) to locate and diagnose the vulnerability. To be concrete, if we want to patch the program, it is vital to hold the source code and non-stripped vulnerable binary at the same time. So it is quite significant to figure out the program configurations of the configuration-related vulnerabilities.


For the developers, the crash report is the most accessable resource, which is collected by several tools (\eg Windows Error Reporting~\cite{debugging}, breakpad~\cite{breakpad}), and records the memory dump with other signals at the crashing time. It is reasonable to figure out the building configurations from code segments extracted from the report. To the best of our knowledge, none of the previous works attempts to infer the building configurations from the crash report to facilitate the reproduction of configuration-related vulnerabilities.


\textbf{Our Approach.} To this end, we present a system called \sys, a practical solution to figure out the configurations in the building process from the code segment of the crash report. The method is beneficial to reproduce the vulnerabilities in the highly configurable programs, even for the further diagnosis and patch process. In this work, we mainly focus on inferring the provenance of \textbf{\textit{building configurations}} (that is how the binary is compiled with specific \textbf{\textit{compilation options}} and \textbf{\textit{program configurations}}) from the crash report.
To consolidate our work, we conduct one empirical analysis of how the different building configurations affect the similarity of generated binaries.


To achieve this goal, we present a comparison-based strategy to extract the configurations from the code segment of the crash report (called \textit{crash report binary} later). Firstly, by recursively generating similar binaries with the crash report binary, we efficiently and precisely search the most possible provenance of \textbf{\textit{compilation options}}. To this step, we hold the crash report binary and the generated binary with the highest similarity on hand. Then, we capture the difference between crash report binary and generated binary, and map it by the self-defined features with the source code to infer the provenance of \textbf{\textit{program configuration(s)}}. Finally, if we successfully get the correct building configurations, we can reproduce the failure, and the program can be carefully diagnosed and patched without any missing information.

To demonstrate the utility of \sys, it is evaluated on a set of 21 real-world configuration-related vulnerabilities via 4 well-known programs. Our experiment shows that \sys could effectively figure out the building configurations needed to generate the vulnerable binary and reproduce the failure, with only 2 cases failed with limited information.

\textbf{Contributions.} The contributions are as follows:

\begin{packed_itemize}
    \item We conduct an investigation on how the different building configurations influence the generated binaries on binary similarity. Based on the discovery, we propose a comparison-based strategy to infer the provenance of building configurations.
    
    \item We implement a prototype called \sys on Linux to facilitate software developers and security analysts in reproducing the configuration-related vulnerabilities.
    
    \item We demonstrate the effectiveness of the system on 21 real-world vulnerabilities. 19 vulnerabilities are successfully reproduced, and the output results are available for further applications, such as root cause diagnosis, vulnerability patching.
\end{packed_itemize}


%% file: background.tex
\section{Background and Motivations} 
\label{sec:bg}


\subsection{Variability of Binary Building}

\vspace{-12pt}
\begin{figure}[htbp]
\centering
    \includegraphics[width=8.8cm]{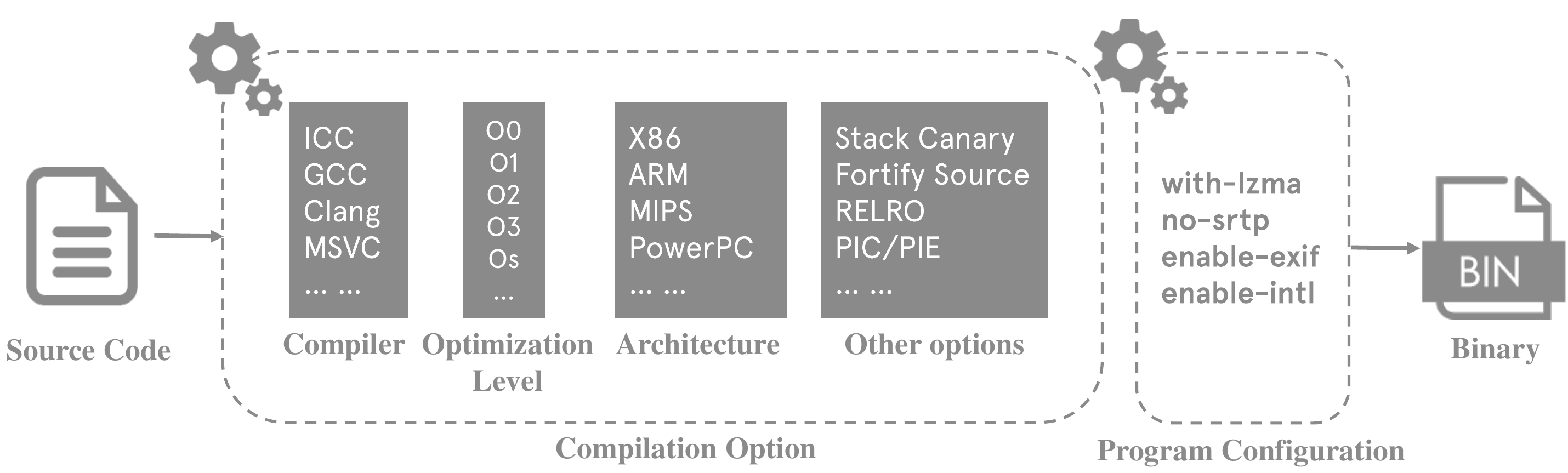}
    \vspace{-10pt}
      \caption{Software Building Process.}
      \label{fig:building}
\end{figure}
\vspace{-8pt}

\textbf{Binary Building.} As we shown in the figure~\ref{fig:building}, when a source code is compiled to a binary code, the final appearance is mainly controlled by two kinds of configurations: \textbf{\textit{compilation option}} and \textbf{\textit{program configuration}}. Compilation options mainly contain the options of compilers, optimization levels, architectures, and security options, while program configurations consist of some options for the specific usages defined by the developers.

\vspace{-4pt}
\begin{table}[htbp]
\centering
\scriptsize
\begin{tabular}{l|ccccc}
\Xhline{1.2pt} Application & libxml2 & OpenSSL & ProFTPD  & PHP & BusyBox \\\Xhline{1.2pt}
\# of Configurations & 41 & 65 & 45 & 131 & 441\\
\Xhline{1.2pt}
\end{tabular}
\vspace{1.8pt}
\caption{Amount of Program Configurations}
\label{tab:configurationNUM}
\end{table}
\vspace{-18pt}

A piece of binary code has a finite combination of different compilation options according to the documents. But the possible combination of program configurations is uncertain. 
To further demonstrate the variability of building a binary, we investigate the amount of program configurations on 5 applications, the result is shown in Table~\ref{tab:configurationNUM}. The number in the table indicates that if an application has $t$ program configurations, it theoretically has $2^t$ candidate combinations (enable or disable for each configuration) of program configurations under a specific compilation option setting, which is an enormous search space. 

\vspace{-8pt}
\begin{figure}[!h]
\centering
\begin{minipage}[t]{0.2\textwidth}
\centering
\includegraphics[width=3.6cm]{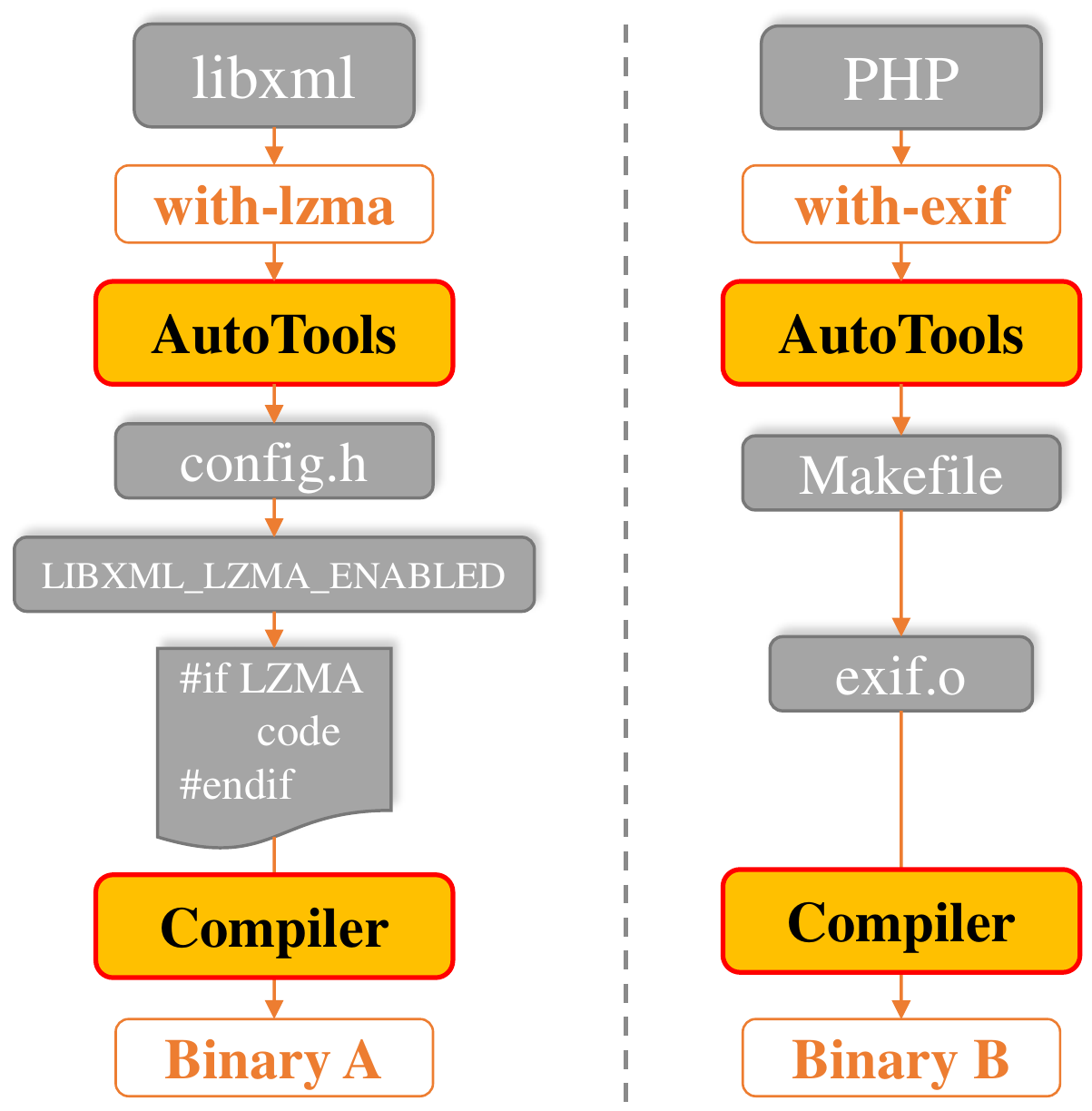}
\caption{Configs and programs.}
      \label{fig:config}
\end{minipage}
\begin{minipage}[t]{0.28\textwidth}
\centering
\includegraphics[width=4.8cm]{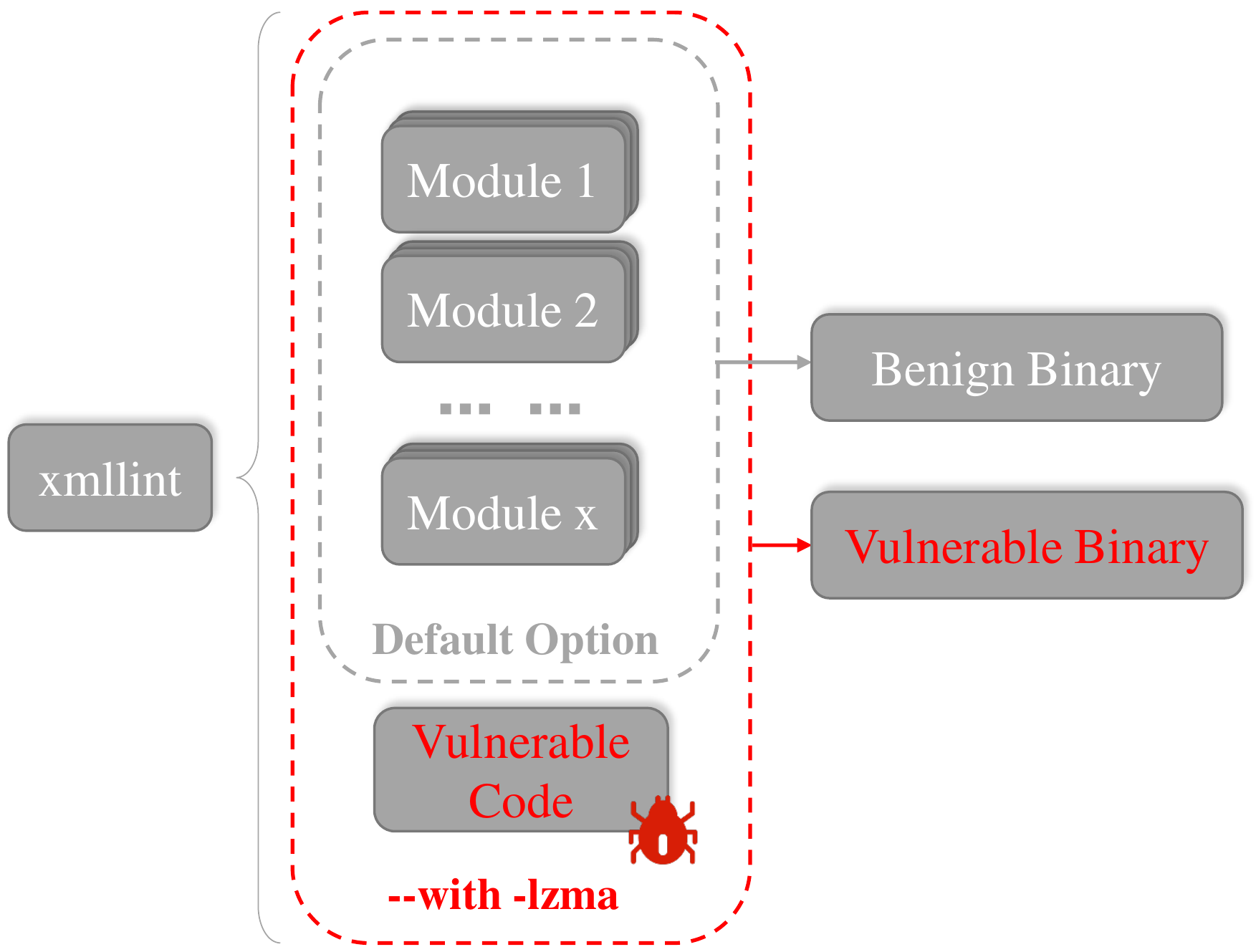}
\caption{Example of CVE-2018-14567}
      \label{fig:MOTIVATION_CASE}
\end{minipage}
\end{figure}
\vspace{-8pt}

\textbf{Configuration-related Vulnerability.}  Configuration-related vulnerabilities only exist under a specific combination of program configurations in a software system.


GNU AutoTools\cite{autotool}, like the GNU compilation system, is a toolset that helps developers adapt source code to various Unix-like systems. Based on the GNU AutoTools compilation system, developers can use Makefile at the file level or at a more fine-grained level to control the content during the compilation process. As shown in the Figure~\ref{fig:config}, libxml uses macros (\eg \textit{LIBXML\_LZMA\_ENABLED}) to control whether the code is compiled into an executable file. When running the config script, it will read the value of the program configurations, to control whether the source code is compiled into the executable file during conditional compilation. As for PHP, the program configurations are realized through file-level (\eg \textit{exif.o})control configuration. It modifies the Makefile to control the compilation process. This mechanism is necessary to support a wide variety of hardware and specific program settings, but it introduces problems in reproducing vulnerabilities in those highly configurable systems.



\textbf{Motivation Example.} Take the following situation as an example. If a specific program configuration controls the vulnerable code, which is missed in the building process, the generated binary is non-vulnerable and the reproduction fails without observing any abnormal behaviors. The vulnerability can not be reproduced without any extra information. Take the case shown in Figure~\ref{fig:MOTIVATION_CASE} as an example. We get a benign binary by compile the \textit{xmllint} with default option, while we get a malicious one with the extra Macro option \textbf{\textit{--with-lzma}}.

So it is obvious to come out a naive solution --- enabling all of the configurations trying to include the vulnerable code. However, it is not practical in the real world after our thorough investigation. On the one hand, some configurations are mutually exclusive. It is impossible to compile all the source code into binary at one time. On the other hand, other configurations in the building process might influence the execution of the vulnerable program, and the generated binary with all possible program configurations will fail to reproduce the target vulnerability. Therefore, we need to treat or include each program configuration conservatively in case they could lead to the failure of reproduction.
\vspace{-10pt}
\begin{figure}[htb]
\centering
    \includegraphics[width=8.5cm]{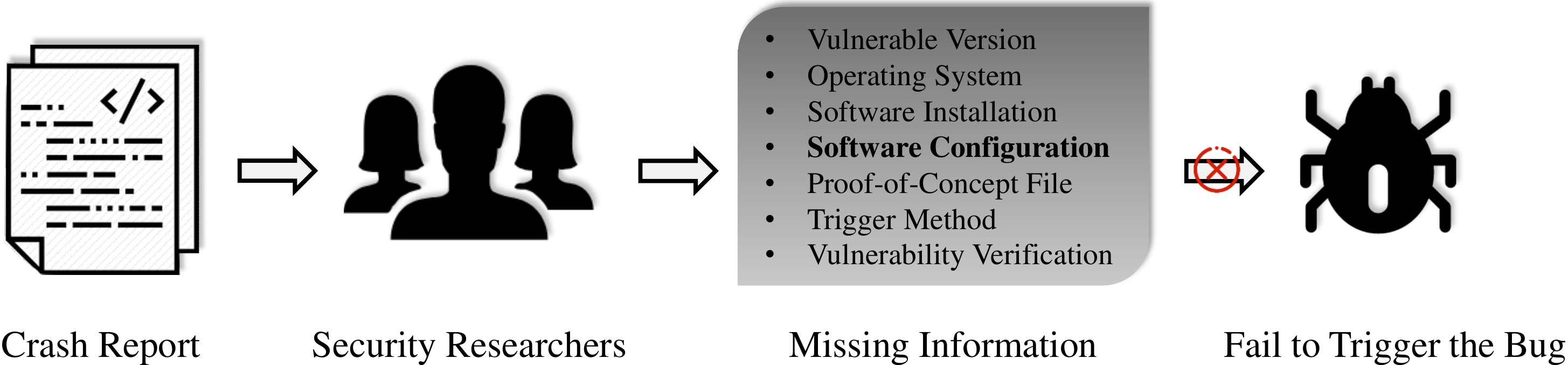}
    \vspace{-10pt}
      \caption{Software Building Process.}
      \label{fig:reproduction}
\end{figure}
\vspace{-10pt}

\subsection{Vulnerability Reproduction}

Nowadays software vendors are increasingly relying on the power of software users to identify security vulnerabilities. Once software developers receive those failure reports from software users, they will reproduce and identify the underlying vulnerabilities promptly. However, previous works have shown failure reports usually lack much key information on vulnerability reproduction~\cite{laukkanen2011survey, Reproduction}. To understand the reproducibility of software vulnerabilities, one study~\cite{Reproduction} conducted an empirical analysis of real-world security vulnerabilities, intending to quantify their reproducibility. The study shows that vulnerability reproduction is difficult and needs intensive manual efforts. And software installation is usually a missing key factor that makes vulnerability reproduction difficult, which is shown in Figure ~\ref{fig:reproduction}. The software installation builds the vulnerable binary and installs it to one specific path. And the key point in this step is how to build the vulnerable binary from source code. That is, in our scenario, it is vital to get the precise program configuration(s).

\vspace{-6pt}
\begin{figure}[!h]
    \centering
    \includegraphics[width=8.8cm]{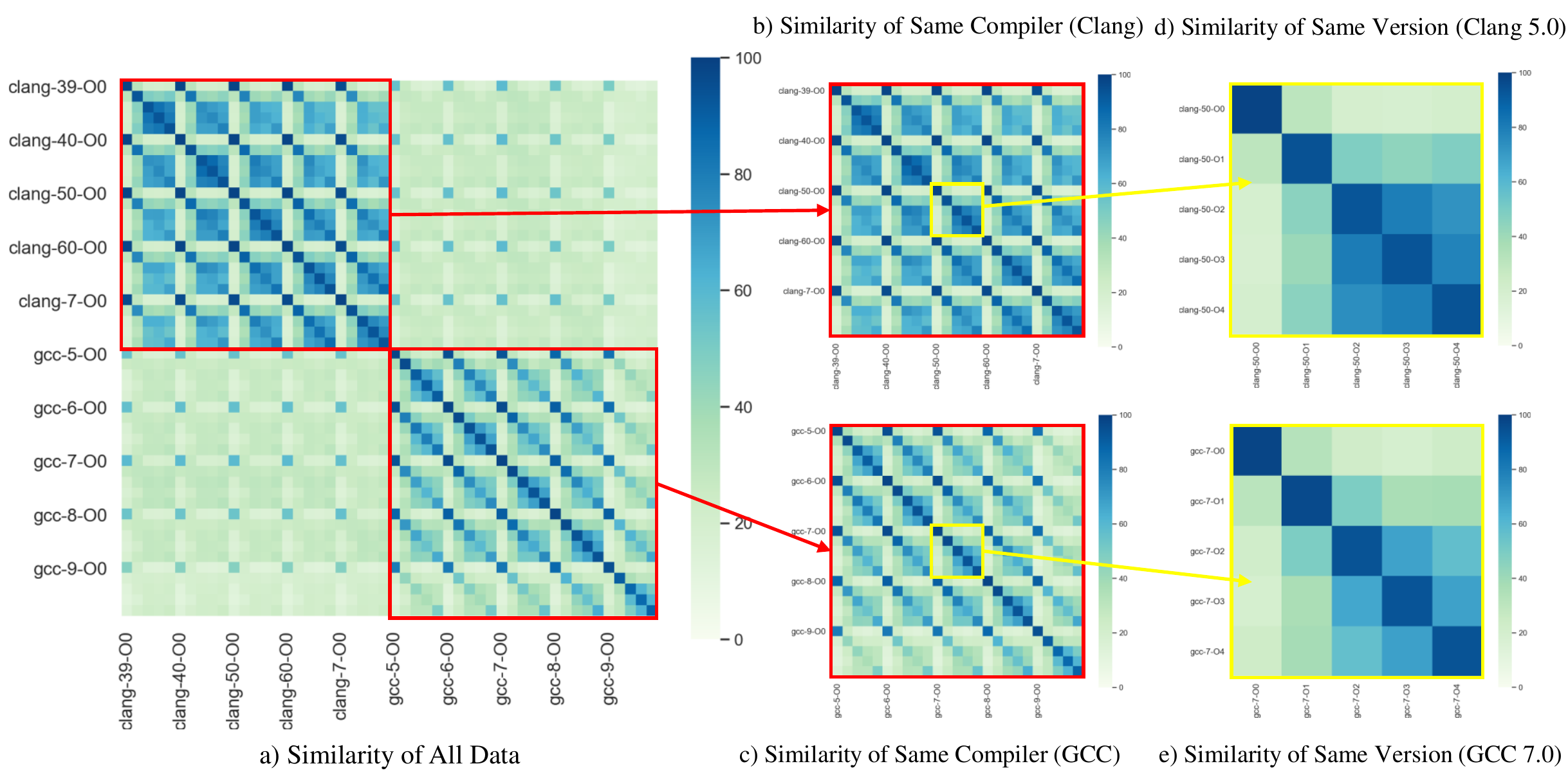}
    \vspace{-10pt}
    \caption{Cross-Compiler, Cross-Optimization Level, Cross- Compiler Version Binary Similarity Comparison by BinDiff }
    \label{fig:similarityanalysis}
\end{figure}


\subsection{Binary Similarity}

\textbf{Configuration Provenance.} The most straight forward method for us is to directly infer the configuration provenance from the crash report binary. Some works has contribute to this field, inferring the compilers\cite{rahimian2015bincomp,massarelli2019investigating}, optimization levels\cite{chen2018himalia,yang2019understand}, even compiler versions\cite{rosenblum2011recovering}. Unfortunately, none of these works can precisely infer all of the compilation options, let alone program configurations. Some machine learning-based methods are even lacking interpretability.

\textbf{Investigation on Binary Size.} To accomplish the mission of inferring the program configuration(s), we need to come out with a new solution. Firstly, we investigate how the specific program configurations influence the binary size.

As shown in Table~\ref{tab:sizeinfluencedbyconfig}, we present the result on our 21 configuration-related vulnerabilities. Columns 1 and 2 denote the CVE (Common Vulnerabilities and Exposures) numbers and the vulnerable related configurations. Column 3 and 4 represent the size of binary without the configuration and with the configuration, in kilobytes (KB). Column 5 denotes the ratio calculated by dividing column 3 by column 4. The table shows that some configurations include more code, while others exclude some code, which can be indicated by the name of configurations and the ratio (whether larger than 1). And the ratio also shows that in most situations, the configurations only take a small part in the whole program. It is hard to come up with an end-to-end method directly inferring the building configurations to overcome the challenge of the unpredictable behavior of configurations and a tiny influence on binary size. So we try to solve the problem by a comparison based method with the help of binary diffing techniques.

\vspace{-8pt}
\begin{table}[htbp]
\centering
\scriptsize
\begin{tabular}{c|ccc|c}
\Xhline{1.2pt}
CVE No.& Configuration & no-Config. & with-config. &  Ratio  \\\Xhline{1.2pt}
CVE-2015-8035  & with-lzma       & 1,263.24       & 1,271.45         & 0.994 \\
CVE-2017-18258 & with-lzma       & 1,265.37       & 1,273.84         & 0.993 \\
CVE-2018-9251  & with-lzma       & 1,265.29       & 1,273.75         & 0.993 \\
CVE-2018-14567 & with-lzma       & 1,265.29       & 1,273.75         & 0.993 \\
CVE-2014-3513  & with-srtp       & 2,679.84       & 2,676.15         & 1.001 \\
CVE-2014-3568  & no-ssl3         & 2,679.84       & 2,676.15         & 1.001 \\
CVE-2014-3569  & no-ssl3         & 2,679.84       & 2,674.96         & 1.002 \\
CVE-2016-6304  & with-ocsp         & 582.09         & 554.1            & 1.051 \\
CVE-2007-1001  & with-gd         & 4,013.64       & 4,328.24         & 0.927 \\
CVE-2016-6294  & enable-intl     & 9,079.28       & 9,484.98         & 0.957 \\
CVE-2016-6297  & enable-zip      & 9,079.28       & 9,179.58         & 0.989 \\
CVE-2019-9025  & enable-mbstring & 11,748.50      & 13,200.21        & 0.890 \\
CVE-2019-9638  & enable-exif     & 9,079.28       & 9,121.79         & 0.995 \\
CVE-2019-9641  & enable-exif     & 9,079.28       & 9,121.79         & 0.995 \\
CVE-2019-9640  & enable-exif     & 9,079.28       & 9,121.79         & 0.995 \\
CVE-2019-9639  & enable-exif     & 9,079.28       & 9,121.79         & 0.995 \\
CVE-2009-3639  & mod\_tls        & 613.07         & 689.14           & 0.890 \\
CVE-2010-4652  & mod\_sql        & 613.07         & 689.14           & 0.890 \\
CVE-2013-4359  & mod\_sftp       & 668.44         & 1,031.25         & 0.648 \\
CVE-2015-3306  & mod\_copy       & 803.79         & 807.58           & 0.995 \\
CVE-2016-3125  & mod\_tls        & 668.44         & 777.68           & 0.860\\\Xhline{1.2pt}
\end{tabular}
\vspace{1.8pt}
\caption{Binary Size Influenced by Configurations}
\label{tab:sizeinfluencedbyconfig}
\end{table}
\vspace{-18pt}

\textbf{Binary Diffing.} The goal of binary comparison is to determine the differences between binaries on the level of syntax, structure, or semantics. In our scenario, we prefer the comparison on the structure of binary code that compares the graph representation of binary code (\eg control flow graphs, call graphs). The structural similarity is more robust to code structure changes and could apply to a large piece of binary code. Binary similarity comparison is a hot top in the field, previous works\cite{duandeepbindiff,liu2018alphadiff,xu2017neuralSIM,zuo2018neuralINNEREYE} equipped with machine learning methods have achieved great performance on the different granularity (\eg binary, function, basic block, etc.). But none of them can meet our need, even the techniques can be used in the situation of cross optimization levels, compiler, architecture, even under obfuscation. They can not help to pinpoint the precise location influenced by the specific program configuration, in other words, the vulnerable part. It leads us to a common but effective method -- recursively generate the binaries to make it more similar to the crash report binary in each step.

\textbf{Commercial Tool BinDiff.}
To this end, we leverage commercial tool BinDiff~\cite{bindiff} as feedback in each recursive step. It is a good diffing tool in representing the structural similarity, which highly relies on the result of IDA Pro. It is related to two research works~\cite{Flake2004StructuralCO, Dullien2005GraphbasedCO}. In 2004, Halvar Flake developed a graph-based binary diffing approach to compare two different binaries generated from the same source code from the perspective of binary code structure~\cite{Flake2004StructuralCO}. To be detailed, it heuristically constructs a subgraph isomorphism on the call graph between two sets of functions in different versions of the same executable binary. To extend this approach into a basic block level, Thomas Dullien~\etal introduced Small Primes Product (SPP) to determine similar basic blocks in the matched functions~\cite{Dullien2005GraphbasedCO}. Therefore, BinDiff could summarize similar functions and the differences between those matched functions, which can effectively assist our method.

%% file: analysis.tex
\section{Analysis on Binary Comparison} 
\label{sec:analysis}


Like the aforementioned, the specific value of program configurations is hard to be extracted from crash report binary with scattered information. So it need to be split the goal into two stages: \textbf{compilation option inference stage} and \textbf{program configuration inference stage}. The reason for firstly inferring the compilation options is because the features of them are distributed globally across the binary, and the information of program configurations need to be matched more precisely. Once the compilation options are confirmed, \sys can concentrate on finding the configurations without being distracted by other information.

In this section, an empirical study is conducted on the binary similarity measured by BinDiff, to investigate how the binaries behave differently that are compiled from the same source code but with different compilation options. The investigation result can guide us to figure out the suspected \textit{compilation options} by leveraging the feedback of BinDiff. 

To scope the field, \sys is experimented on 2 compilers (GCC and Clang), 5 optimization levels (O0, O1, O2, O3, Os separately), and 5 versions for each compiler (5.0, 6.0, 7.0, 8.0, 9.0 for GCC, 3.9, 4.0, 5.0, 6.0, 7.0 for Clang). Under this setting, a piece of source code has 50 ($2 \times 5 \times 5$) possible combinations of compilation options. Figure~\ref{fig:similarityanalysis} presents the cross-comparison between 50 sets of binaries compiled with different compilation options, measured by BinDiff. Figure~\ref{fig:similarityanalysis} a) shows the full scene of the result. Due to the limited space, part of the description of the grids is omitted. From the top row to the bottom, the grid separately denotes Clang-39-O0, Clang-39-01, Clang-39-O2, Clang-39-O3, Clang-39-Os, Clang-40-O0, ..., and so on. The color of the grid represents the grade of similarity between the two binaries. The darker, the more similar. For instance, the grid in row 1 column 2 denotes the similarity between the binaries compiled with Clang-39-O0 and Clang-39-O1, which are not that similar. As for the sub-figure b) and c), it is easier to capture the trend of similarity of the same compiler. And the sub-figure d) and e) that shows the trend in the same version but with different optimization levels.

According to the figures, the following findings inspire the further procedure of inferring compilation options:


\begin{packed_itemize}
\item \textbf{Findings for Sub-Figure a)}
\ding{182} Binaries compiled with -O0 behave quite differently from -O1, -O2, -O3, and -Os, no matter which compiler. 
\ding{183} In the vertical comparison, binaries compiled by the same compiler has a higher similarity under the same optimization level.

\item \textbf{Findings for Sub-Figure b) and c)}

\ding{184} Neighboring compiler versions (\ie GCC-5 and GCC-7 are the neighboring compiler versions of GCC-6) process more similar than others, which will slightly influence how the optimization level affect the binary.
\ding{185} Binaries compiled with the same optimization level behave more similar from the same compiler, no matter which version.

\item \textbf{Findings for Sub-Figure d) and e)}
\ding{186} Neighboring optimization levels (\ie -O1 and -O3 are the neighboring optimization levels of -O2) make the binary more similar, but only ranging from -O0 to -O3. -Os behave more similar with -O2 rather than -O3.


\end{packed_itemize}

%% file: definition.tex
\section{Problem Statement}
\label{sec:problem}

In our work, we aim to facilitate the reproduction of configuration-related vulnerabilities. After the investigation and analysis, it is the key point that figuring out the building configurations from the crash report binary -- \textbf{compilation options} and \textbf{program configurations}. After we get the configurations, we can generate a piece of binary highly similar to the crash report binary, which can be leveraged for the downstream applications, such as root diagnosis, vulnerability patch, etc. To make the solution more concrete, we define our problem as follows.

\vspace{-8pt}
\begin{figure}[!h]
    \centering
    \includegraphics[width=6.5cm]{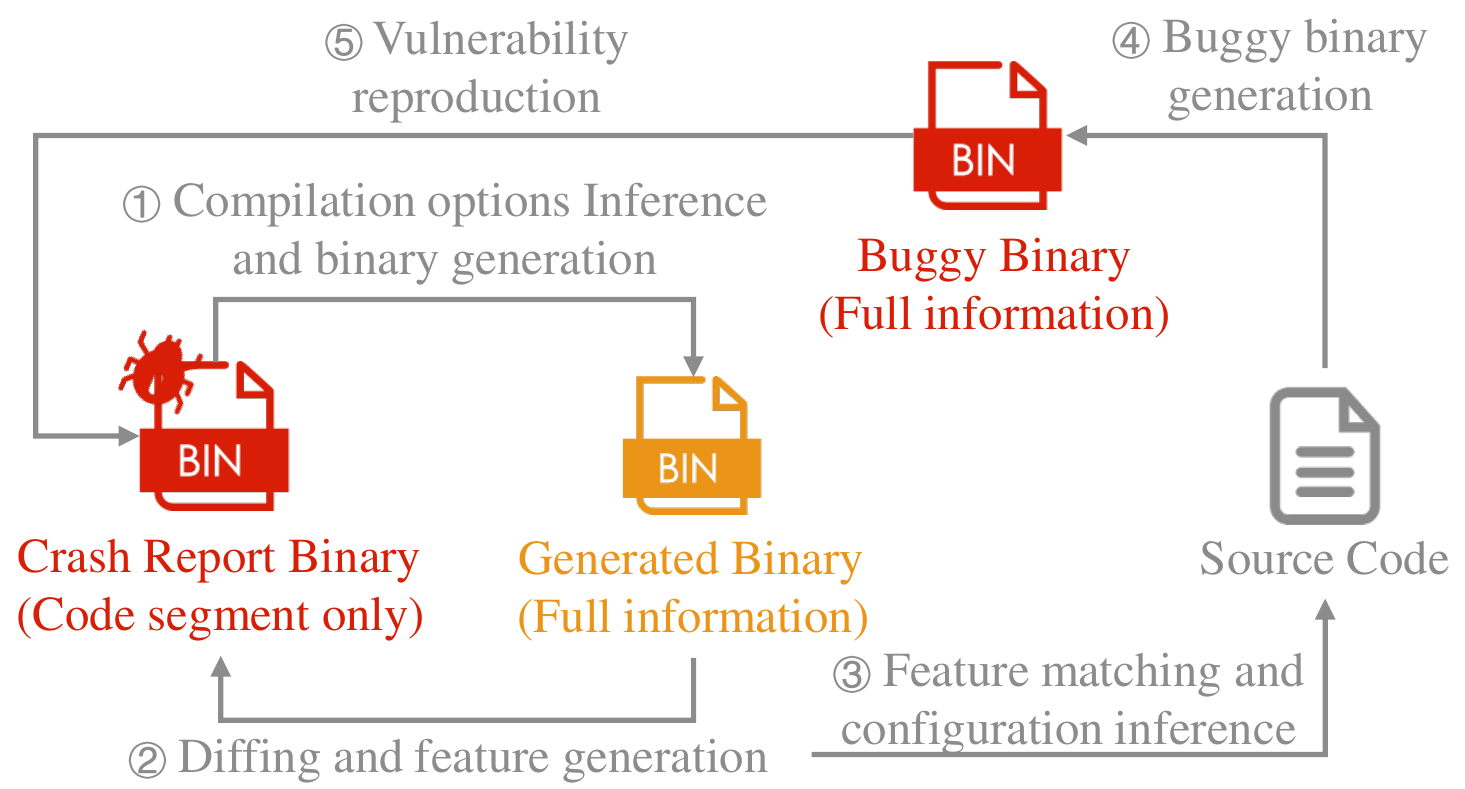}
    \vspace{-10pt}
    \caption{Brief Workflow according to \sys}
    \label{fig:problemstatement}
\end{figure}
\vspace{-8pt}

\subsection{Problem Definition}
\begin{packed_itemize}
    \item \textbf{Input:} crash report binary $crB_t$, open source program $S_x$
    \item \textbf{Intermediate Result: } generated binary $gB_s$
    \item \textbf{Output:} buggy binary $bB_n$
\end{packed_itemize}

According to the brief workflow shown in Figure~\ref{fig:problemstatement}, \sys leverages generated binary $gB_s$ as an intermediate result to bridge the crash report binary $crB_t$ and open source program $S_x$, and finally generates the buggy binary $bB_n$.

\ding{172} Given a piece of crash report binary $crB_t$, we firstly infer the provenance of compiler $C_i$, optimization level $OL_j$ and compiler version $V_k$ by the recursive generation step as $R(crB_t) = \{C_i, OL_j, V_k\}$, where in this work C = \{GCC, Clang\}, OL = \{O0, O1, O2, O3, Os\}, V = \{GCC 5.0, GCC 6.0, GCC7.0, GCC 8.0, GCC 9.0, Clang 3.9, Clang 4.0, Clang 5.0, Clang6.0, Clang 7.0\}.

With the information of compilation options above, we can generate a binary $gB_s=G(S_x, C_i, OL_j, V_k)$ from the open source program $S_x$. The similarity of two binaries $Sim(crB_t, gB_s)$ exceeds the preset threshold, which indicates that they are exactly the same except for the program configuration(s). \ding{173} Finally, with the difference $Diff(crB_t, gB_s)$ between the binaries and the features $F(S_x)$ captured from the source code, \ding{174} we are able to find the relationships and infer the setting of configuration(s) $Cinfig_y=\{config_1 = True, config_2 = Flase, ...\}$.

Till this step, we have all of the building configurations and the source code on hold, \ding{175} so that we can generate the buggy binary $bB_n=G(S_x, C_i, OL_j, V_k, Config_y)$. \ding{176} And it is natural to reproduce the vulnerability with the buggy binary $bB_n$, which can also assist most of the downstream applications in the security field.





%% file: overview.tex
\section{System Design} 
\label{sec:overview}

\begin{figure*}[htbp]
    \centering
    \includegraphics[width=16cm]{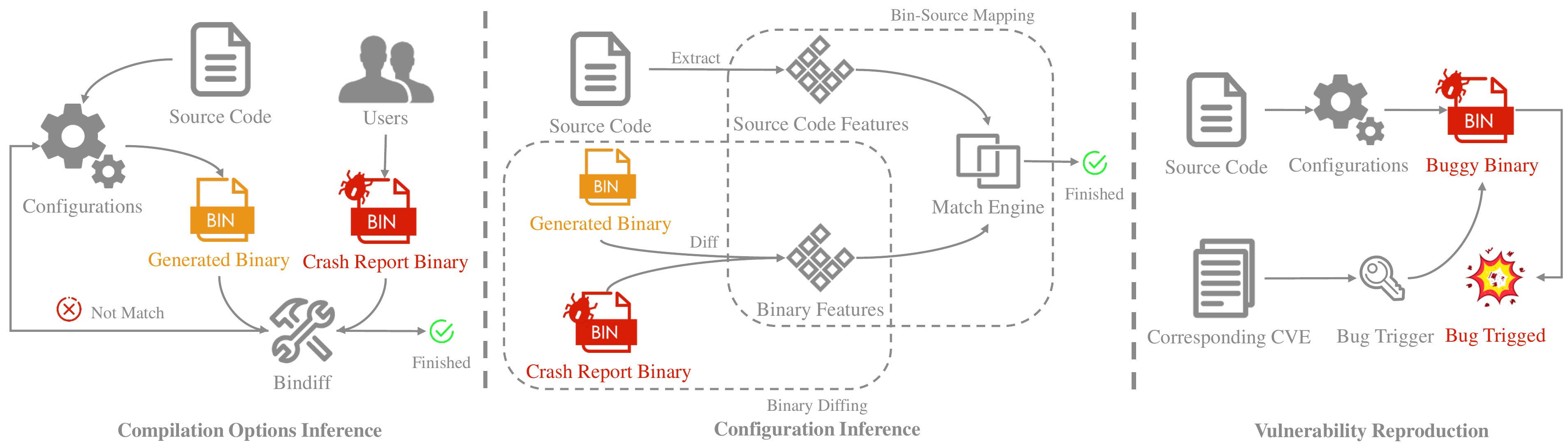}
    \vspace{-10pt}
    \caption{Workflow of building configuration inference}
    \label{fig:workflow}
\end{figure*}

Figure~\ref{fig:workflow} delineates the architecture of \sys. The whole system contains three major components: \textbf{Compilation Options Inference}, \textbf{Configuration Inference} and \textbf{Vulnerability Reproduction}.

To summarize, our work focus on facilitating the reproduction of configuration-related vulnerabilities by hierarchically inferring the compilation options and program configurations.

\textbf{Compilation Options Inference.} Firstly, with the source code and the code segment of the crash report (called \textit{Crash Report Binary} in the figure) on hold, we recursively generate the binaries to compare with the crash report binary. By leveraging the commercial tool BinDiff, we are able to measure the similarity of the generated binary and the crash report binary. When the similarity reaches a certain threshold, we consider the provenance of the compilation options of the current generated binary is the same as the crash report binary.

\textbf{Configuration Inference.} In this stage, we regard the only difference between the current generated binary and the crash report binary is program configuration. We extract the binary level features by diffing the two binaries, while the source level features are extracted from the source code. To find the candidate program configurations by mapping the features, we employ a customized matching engine. Finally, with the help of the solver, we can eventually target the specific program configuration(s). 

\textbf{Vulnerability Reproduction.} Following the former stages, we are holding the building configurations of the vulnerable binary that can help generate the buggy binary. According to the report, we are able to reproduce the failure. Furthermore, we narrow the vulnerability to a specific and limited scope, so we can precisely apply it to the diagnose or the patch procedure.

%% file: design.tex
\subsection{Compilation Options Inference}



To reproduce the vulnerability, we need to generate a binary that is the most similar to the one in the crash report.
We study the software building process which translates human-readable source code to one executable binary. According to our survey, we find that program configurations slightly influence the binary, while compilation options globally have an impact on the executable. To precisely get the building configurations and reproduce the configuration-related vulnerabilities, we firstly determine the compilation options. Based on this, we are easier to figure out the program configurations.

In section~\ref{sec:analysis}, we summarize some findings from the analysis results. Based on them, we search the compilation options by the following steps.

\ding{172}\textbf{Whether -O0 or NOT?} Firstly, we compile the source code by default version of GCC (or Clang) separately with -O0 and -O2. By comparing with the crash report binary, the similarity produced by BinDiff will tell whether the provenance of the optimization level is -O0.

\ding{173}\textbf{Which Compiler?} Then, we compile the code separately by GCC and Clang with the same optimization level produced by the former step. The one that has a higher similarity with the crash report binary can tell the provenance of the compiler.

\ding{174}\textbf{Optimization Level Confirmation.} If the provenance of the optimization level is -O0, we do not have to go through this step. Conversely, we generate the binary by the same compiler and its version with -O1, -O2, O3 to confirm the right optimization level by the highest similarity.

\ding{175}\textbf{Compiler Version Confirmation.} Last but not least, we confirm the version of the compiler. By comparing the similarity of neighboring versions, we can finally get all of the compilation options.







Leading by the steps above, we are feasible to figure out the specific compilation options from the large search space. Due to the limitation that the mechanism of BinDiff is rule-based, we can not guarantee that the generated binary has the full similarity with the crash report binary. But the binary with a relatively high similarity is prepared for the next stage.

\begin{figure}[htbp]
\centering
    \includegraphics[width=8.5cm]{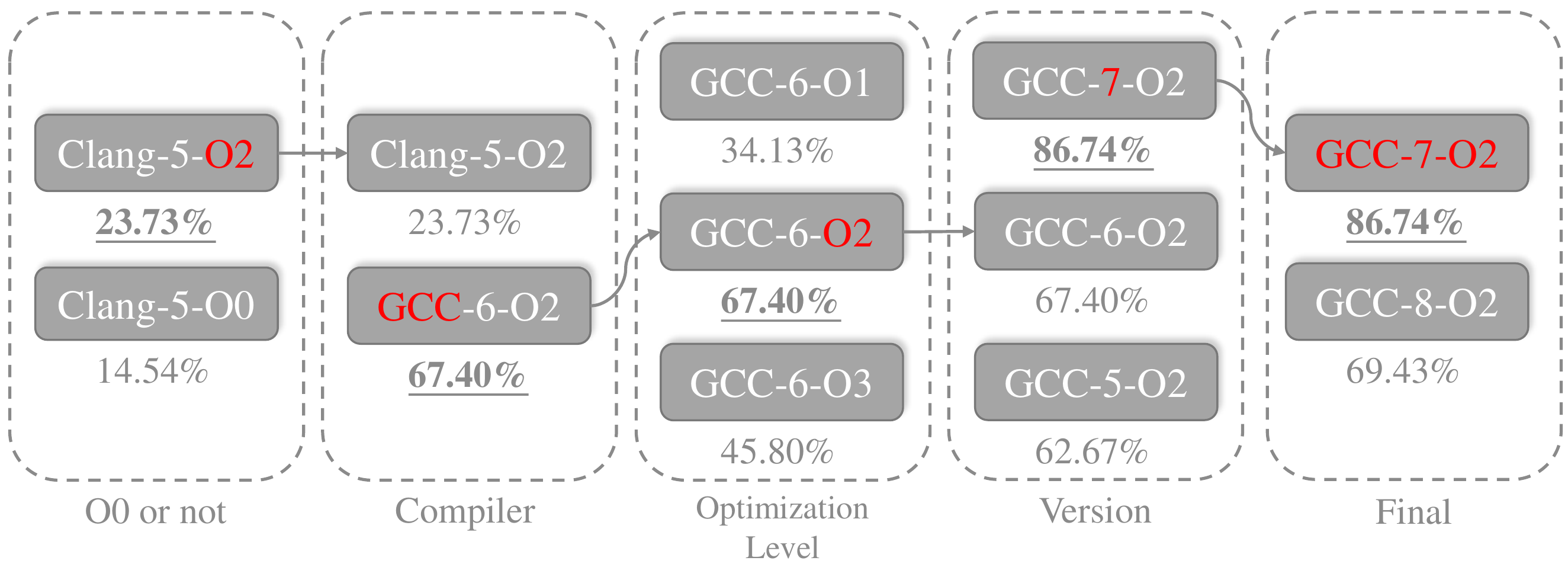}
    \vspace{-8pt}
      \caption{An Show Case of Compilation Options Inference}
      \label{fig:compilationinference}
    \vspace{-10pt}
\end{figure}

We take an example to show the procedure. As shown in Figure~\ref{fig:compilationinference}, we present the process of how to figure out the provenance of compilation options of \textit{xmllint} compiled with compilation options \textit{GCC-7-O2}. All the percentages in the figure are the similarity comparison result between the generated binary and the crash report binary produced by BinDiff.

In the first step, we confirm \ding{172}whether it is compiled with -O0. By compiling the source code with the same default version of Clang (randomly choose the compiler) but different optimization levels, we can make sure the optimization level is NOT -O0. Then, we compare the generated binary compiled with \ding{173}different default compiler. The result indicated the provenance of the compiler is GCC. Next, by generating the binary with neighboring optimization levels, we can confirm the \ding{174}exact optimization level -O2. Finally, by listing all of the candidate compiler versions, we can figure out the \ding{175}specific compiler version. It takes 8 times (we do not repeatedly compare binaries with the same compilation options) of comparison to figure out the options. And the process of compilation options inference meets an end till now. 

\subsection{Binary Diffing}

This stage mainly contains three parts. According to Figure~\ref{fig:workflow}, we need go through \textbf{Binary Diffing}, \textbf{Source Code Extraction} and \textbf{Bin-Source Mapping} to figure out the specific program configuration(s).

To map the differences from binary to source code, we firstly leverage IDA Pro~\cite{disassembler2010debugger} and Bindiff~\cite{bindiff} to identify identical and similar functions between the generated binary and the crash report binary. So the rest of the parts are the candidates controlled by the program configurations.

Due to the behavior of program configurations is unpredictable, some are '-enable' and some are '-disable', so the size of the relationship between the generated binary and the crash report binary is uncertain, which is presented in Table~\ref{tab:sizeinfluencedbyconfig}. To confirm our observation, we investigate our data set.
The result shows three kinds of situation as follows:

\begin{itemize}
    \item Some functions only appear in the generated binary;
    \item Some functions only appear in the crash report binary;
    \item Functions appear in both of them.
\end{itemize}

Because the size of the crash report is limited to compress the content, the crash report binary always does not contain symbolic information. Therefore, we need to position the differences as much as possible in the generated binary. According to the situations above, we give different solutions.

For the first situation, we realize those functions are not compiled into the vulnerable binary or disabled by the configurations. In the subsequent Bin-Source mapping step, the constraints of these types of functions can be obtained, and only the logical relationship 'and' is needed as the connection. To obtain the final constraints, we only need to reverse them. However, conflicts may arise during the procedure. Our strategy is to discard conflict functions. On the one hand, discarding them has little impact on the result. On the other hand, conflicts are very rare and almost all of them appear in function re-definitions. If we do not find any configurations, it may be a false alarm of BinDiff.


In the second situation, it is exactly opposite to the first one.  The process is similar to the first situation. The only difference is that we do not need to reverse the constraint. But we need to extract features separately from the crash report binary and source code due to the missing symbolic information of the crash report binary. After feature matching, a set of constraints is obtained, and the provenance of program configuration can be extracted by the solver.

For the last situation, we continue to detect and highlight binary differences between two variants of the same function. We use some fine-grained methods to map those binary differences to the corresponding source code. If the functions are exactly the same, we can easily extract and map the constraints. If the functions are partially matched the rest partially matched functions, it is due to the different value of Macros inside the function, which leads to the difference of internal structures.

To better analyze the result, we take a piece of binary code from \textit{XMLlint} and visualize part of the structure of the generated binary and crash report binary with the help of IDA Pro and BinDiff.


\begin{figure}[htbp]
\centering
    \includegraphics[width=8.5cm]{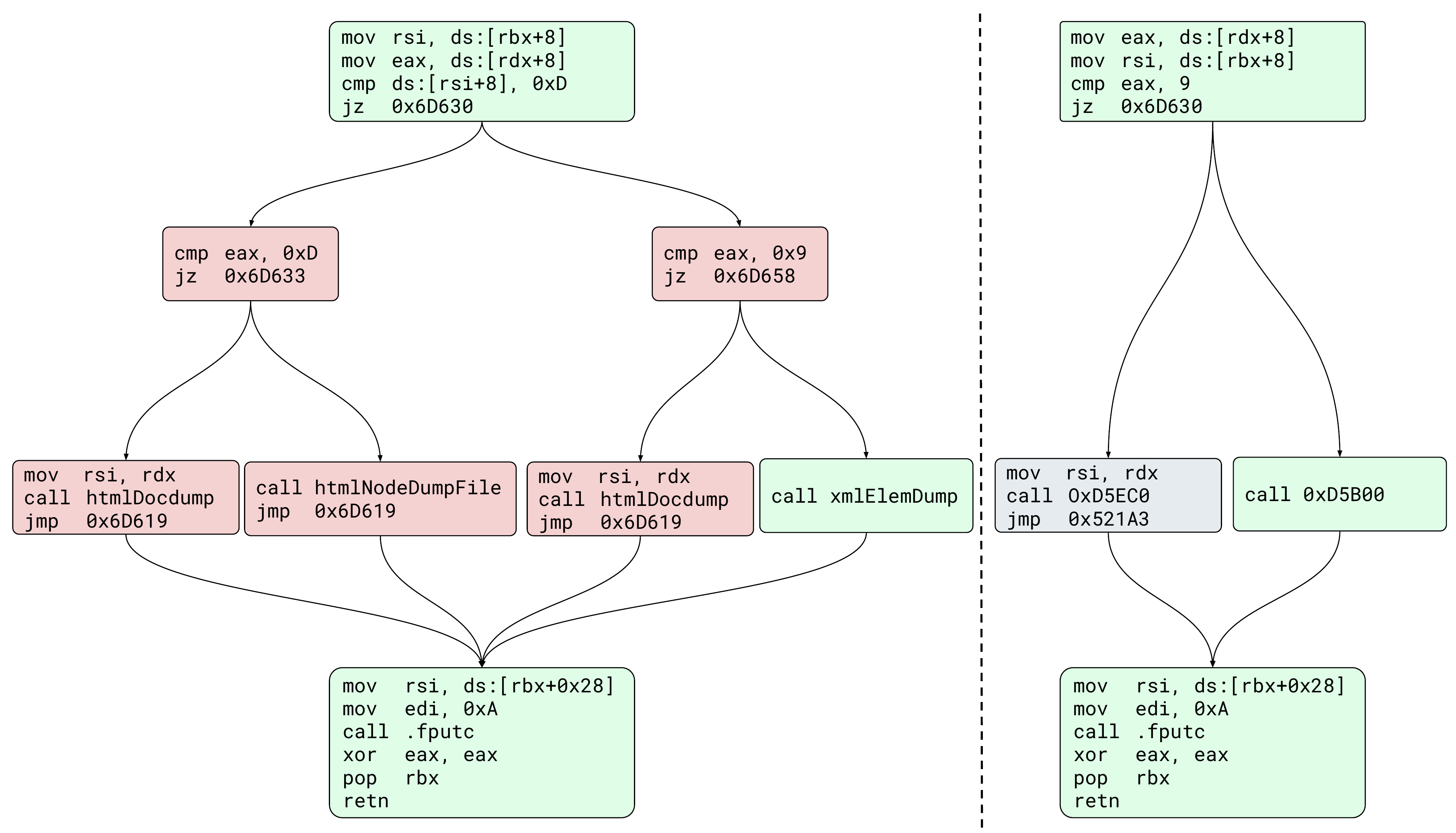}
    \vspace{-8pt}
      \caption{Binary Diffing Example of Generated Binary and Crash Report Binary}
      \label{fig:bindiffexample}

\end{figure}

As shown in Figure~\ref{fig:bindiffexample}, the green parts are the base blocks that exist in both of the binaries, the red parts are the base blocks that only exist in the generated binary, and the gray part is the base block that only exists in the crash report binary.

To better understand how these differences arise, we find out the corresponding source code which is shown as follows. According to the code, it is obvious that different values of the Macro \textit{LIBXML\_HTML\_ENABLED} will control different codes to be compiled into binary. It is worth noting that the code in line 2428-2431 is exactly the same as the code in line 2434-2434. When the Macro \textit{LIBXML\_HTML\_ENABLED} is set as 0, the compiler will optimize the code, and only one copy of the code will be reserved, resulting in the binary on the right side is less than the one on the left according to Figure~\ref{fig:bindiffexample}. It is quite challenging that the diffing result is not only controlled by the program configurations but sometimes also by compiler and optimization levels.

\lstset{
    basicstyle=\scriptsize\ttfamily,
    rulesepcolor= \color{gray}, 
    breaklines=true,  
    commentstyle=\color{gray}, 
    frame=shadowbox
    }

    \begin{lstlisting}
2421     if (ctxt->doc->type == XML_HTML_DOCUMENT_NODE) { 
2422 #ifdef LIBXML_HTML_ENABLED 
2423         if (node->type == XML_HTML_DOCUMENT_NODE) 
2424             htmlDocDump(ctxt->output, (htmlDocPtr) node); 
2425         else 
2426             htmlNodeDumpFile(ctxt->output, ctxt->doc, node); 
2427 #else 
2428         if (node->type == XML_DOCUMENT_NODE) 
2429             xmlDocDump(ctxt->output, (xmlDocPtr) node); 
2430         else 
2431             xmlElemDump(ctxt->output, ctxt->doc, node); 
2432 #endif /* LIBXML_HTML_ENABLED */ 
2433     } else { 
2434         if (node->type == XML_DOCUMENT_NODE) 
2435             xmlDocDump(ctxt->output, (xmlDocPtr) node); 
2436         else 
2437             xmlElemDump(ctxt->output, ctxt->doc, node); 
2438     } 
2439     fprintf(ctxt->output, "\n"); 
2440     return (0); 
2441 }

    \end{lstlisting}



Till now, we have the rough constraints extracted by the diffing result from generated binary and crash report binary. We still need to map them to the source code, so that we can figure out the specific program configurations (\ie configuration files, macros).







\subsection{Feature Extraction}



The exact correspondence between binary and source code is an open and difficult problem in the security field. So we come up with the second-best solution. Based on the extraction result of the binary diffing, we can get the corresponding source code range. However, it is impossible to determine which macro-related code fragment in the source code appears in the crash report binary. In addition, when there are more than one macros in the source code range, it is challenging to directly infer the specific configurations from the generated binary. Therefore, we separately extract the source code fragment features related to the macro and search for the source code feature in the binary to determine which code fragment is compiled into the binary.
To tackle the problem, we need to find out some \emph{features} to assist our work. The ideal feature needs to have the following criterion.

\textit{\textbf{Unique.}}
The feature should not appear multiple times within the specified range, otherwise, it cannot be uniquely identified. The range is adjustable to the mapping space, because we may need coarse-grained function-level matching and fine-grained intra-function matching for feature matching. Intuitively, if we need to match a certain block of code in a function, then the range is limited in the function. However, if we need to match a function in the whole binary, then the range is the entire binary. In case a feature is not unique, we will not be able to accurately determine whether the code segment exists.

\textit{\textbf{Stable.} }
The features should be stable enough which cannot be affected by the different compilation options from source code to binary. In other words, these features do not only exist in the source code but also persist in the binary. This requires that the features should not be too complicated, otherwise it may be changed after passing through the compiler and cause mismatching.
To meet the requirements, we select 3 kinds of features: \textit{string}, \textit{constant}, \textit{function call}. All of them appear in both source code and binary code, which can serve as anchor points for matching.


\subsection{Feature Generator}



With the aforementioned features, we add some structural features \textit{if}, \textit{condition}, \textit{then} branch, and \textit{else} branch. When combined with structural features, we can make the semantic features more unique to improve efficiency. 

\textbf{Source Code Feature Generator.}
TypeChef has been designed to check for incompatible types and developer errors in untested configuration combinations. So we leverage TypeChef to generate a macro AST, called VAST, and extract source code features from it. Each extracted feature has the constraints of the features.
\textbf{Binary Feature Generator.}
We leverage IDA Pro to extract the features from binaries.
If more accurate information is needed, we will use some procedural analysis techniques, such as \textit{binary function parameters}, \textit{data streams} to assist us in more precise binary matching.

\subsection{Feature Matching Engine}



In the feature matching stage, we start with a coarse-grained matching strategy for efficiency. When it fails, we complement it with a fine-grained matching strategy.

\textbf{Coarse-grained Matching.}
In the coarse-grained matching, we mainly consider the simple features, \textit{string}, \textit{constant} and \textit{function call}. For instance, we directly search a string in the source code to see whether the corresponding string information exists in the binary function. If it exists, the match succeeds. If it fails, we select other features to match again and come out with the confidence of the feature identification. Assuming that none of the features are matched successfully, it can be considered that the code does not exist in the binary. Furthermore, we can use this strategy to match the function level. However, this matching strategy is relatively simple and rough, and some of the segments can not be successfully matched.

\begin{figure}[htbp]
    \centering
    \includegraphics[width=8.5cm]{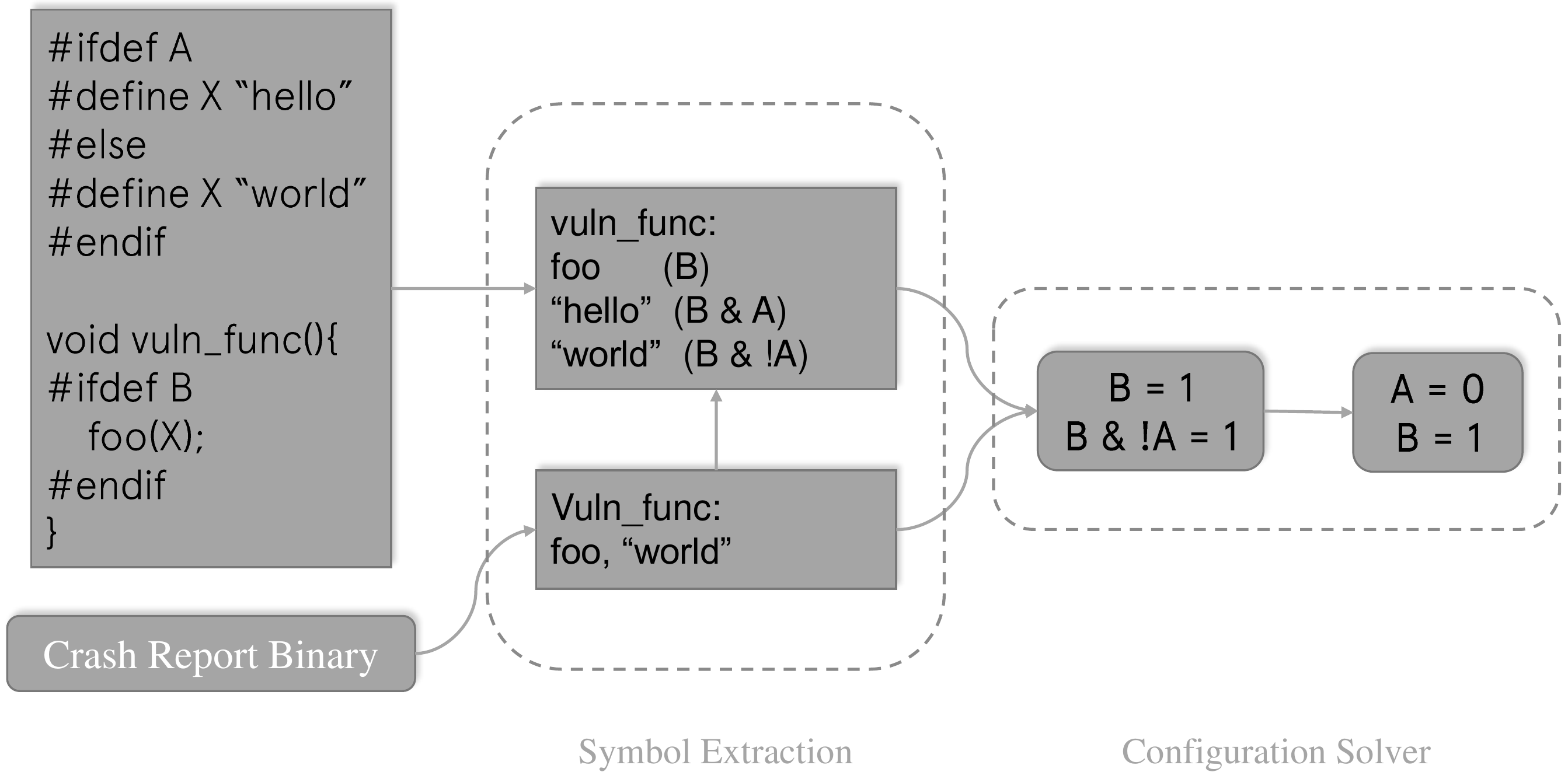}
    \caption{Example for the Solver and Matching Engine}
    \label{fig:solver_example}
\end{figure}

\textbf{Fine-grained Matching.}
In case all the rules of coarse-grained matching fail, we will turn to use a fine-grained matching strategy. Since some features can not be a matching anchor alone, so we combine them with extra structural features (e.g., branch structure, loop structure) to become a more unique identifier. Such as, \textit{if} statements in source code correspond to the instruction \textit{cmp} in binary code. 
As shown in Figure \ref{fig:solver_example}, if the function \textit{foo} is considered alone, once the function call \textit{foo} can be found more than once in the binary, so it is not certain whether the macro-related code is compiled into the binary. But when the combined features are introduced, considering the features of the \textit{if} structure, the constraint can be extracted in the comparison condition, and the function call \textit{foo} can be extracted in the statement block. So when we leverage the context information (i.e., structural feature), the combined information can be used as the smallest matching unit to accurately match the functions affected by the macro.


\vspace{-8pt}
\vspace{-8pt}

\subsection{Solver}



By feature extraction and feature matching, we can confirm whether a code segment controlled by program configuration(s) exists in the target binary (generated binary or crash report binary). Furthermore, we need to figure out which program configuration controls the code segment.
To infer the precise configuration(s) of the program, we speculate the specific values of the program configurations according to the result of feature matching and the existing condition of the source code from VAST. Take the case in Figure\ref{fig:solver_example} as an example, feature matching can find the string \textit{world} in the binary matches the string in the source code. And there is a function call \textit{foo} in the binary, with VAST we can get the condition of \textit{foo}'s existence is $B$, and the existence condition of the string \textit{world} is $\neg A$. Combining the analysis and matching results of the source code, a satisfactory value of macros $A$ and $B$ can be obtained by the constraint solver.

\vspace{-8pt}
\lstset{
    basicstyle=\scriptsize\ttfamily,
    rulesepcolor= \color{gray}, 
    breaklines=true,  
    commentstyle=\color{gray}, 
    frame=shadowbox
    }

    \begin{lstlisting}[caption={Macro Expression of LibPNG}]
1 defined(PNG_FLOATING_POINT_SUPPORTED) &&
2!defined(PNG_FIXED_POINT_MACRO_SUPPORTED) &&
3(defined(PNG_gAMA_SUPPORTED)
4 defined(PNG_cHRM_SUPPORTED)
5 defined(PNG_sCAL_sUPPORTED)
6 defined(PNG_REAd_BACKGroUNd SUPPORTED)
7 defined(PNG_READ_RGB_TO_GRAY_SUPPORTED))
8(defined(PNG_sCAL_SUPPORTED) &&
9 defined(PNG_FLOATING_ARITHMETIC_SUPPORTED))
    \end{lstlisting}

However, the situation is not simple in the real world. We find that to speed up the compilation process, if the value of the expression is \textit{FALSE}, all of the code controlled by the expression will be skipped without analysis. For instance, if any of the macros in LibPNG shown as above is not defined, everything under its control will be excluded from the compilation process.

To tackle this problem, we design an expression analyzer to analyze macro expressions. We employ \textit{Z3} solver to generate an AST (Abstract Syntax Tree) of macro expressions by lexical analysis. With the AST on hold, we are able to figure out the logical relationships between macro expressions. When multiple logical relations are consistent with the case, we can simply merge some conditions, and come out with one satisfied solution. We present a solving result of macro expressions in LibPNG as follows.

    \lstset{
    basicstyle=\scriptsize\ttfamily,
    rulesepcolor= \color{gray}, 
    breaklines=true,  
    commentstyle=\color{gray}, 
    frame=shadowbox
    }
    
\vspace{-8pt}
    \begin{lstlisting}[caption={Analysis Result of Macro Expression}]
1 [PNG_FLOATING_POINT_SUPPORTED = False,
2 PNG_gAMA_SUPPORTED = False,
3 PNG_sCAL_sUPPORTED = False,
4 PNG_cHRM_SUPPORTED = False,
5 PNG_REAd_BACKGroUNd SUPPORTED = True,
6 PNG_FIXED_POINT_MACRO_SUPPORTED = False,
7 PNG_FLOATING_ARITHMETIC_SUPPORTED = True,
8 PNG_FLOATING_POINT_SUPPORTED = True]
    \end{lstlisting}

\subsection{Vulnerability Reproduction}



At this stage, we overcome the challenges mentioned in the previous sections and infer the compilation options and program configurations from the crash report binary. With the source code and the inferred building configurations, we are not only able to reproduce the failure, but also can precisely locate the vulnerability. Furthermore, we can leverage the result to diagnose and patch the vulnerable part.


%% file: eval.tex
\section{Evaluation} 
\label{sec:eval}





\subsection{Procedures and Settings}

Due to the inconsistency of CVE and the bug report~\cite{Reproduction,dong2019towards}, only part of the vulnerabilities is crash reports available. So we simulate the scene to evaluate our work. It will be our future work that facilitating the reproduction of vulnerabilities by leveraging limited and incomplete information.

\begin{table*}[htbp]
\centering
\scriptsize
\begin{tabular}{c|c|ccccc|cc|cc}
\Xhline{1.2pt}
\multicolumn{1}{l|}{}     & \multicolumn{1}{c|}{\multirow{2}{*}{CVE ID}} & \multicolumn{5}{c|}{Compilation Options   Inference}                                                  & \multicolumn{2}{c|}{Configruation   Inference} & \multicolumn{2}{c}{Reproduction} \\
\multicolumn{1}{l|}{}     &    \multicolumn{1}{l|}{}                    & Version & $t_{Build}$ (s) & compilation options & $T_{Infer}$ (\#) & $\beta$ & Config.              & $t_{Extract} (s) $     & Defualt          & Robin         \\\Xhline{1.2pt}
\multirow{4}{*}{libxml}  & CVE-2015-8035        & 2.9.1               & 22.67            & GCC-7-O0            & 5              & 98.22\%              & with-lzma            & 28.96                  & \xmark              & \cmark           \\
                         & CVE-2017-18258       & 2.9.6               & 18.38            & GCC-5-O3            & 7              & 86.53\%              & with-lzma            & 33.32                  & \xmark              & \cmark           \\
                         & CVE-2018-9251        & 2.9.8               & 19.78            & GCC-8-O1            & 7              & 95.33\%              & with-lzma            & 182.17                 & \xmark              & \cmark           \\
                         & CVE-2018-14567       & 2.9.8               & 19.24            & Clang-6-O1          & 7              & 94.60\%              & with-lzma            & 181.37                 & \xmark              & \cmark           \\\hline
\multirow{4}{*}{OpenSSL} & CVE-2014-3513        & 1.0.1i              & 65.28            & GCC-5-03            & 7              & 73.05\%              & with-srtp              & 102.19                 & \cmark              & \cmark           \\
                         & CVE-2014-3568        & 1.0.1i              & 66.54            & GCC-7-O0            & 5              & 98.94\%              & no-ssl3              & 43.68                  & \xmark              & \xmark           \\
                         & CVE-2014-3569        & 1.0.1i              & 67.45            & Clang-6-O1          & 7              & 97.61\%              & no-ssl3              & 44.52                  & \xmark              & \xmark           \\
                         & CVE-2016-6304        & 1.1.0               & 81.16            & GCC-8-02            & 7              & 95.61\%              & with-ocsp              & 193.47                 & \cmark              & \cmark           \\\hline
\multirow{8}{*}{PHP}     & CVE-2007-1001        & 5.2.0               & 109.09           & Clang-6-O1          & 7              & 87.89\%              & with-gd              & 184.33                 & \xmark              & \cmark           \\
                         & CVE-2016-6294        & 7.0.0               & 308.3            & GCC-5-O2            & 8              & 87.50\%              & enable-intl          & 1712.87                & \xmark              & \cmark           \\
                         & CVE-2016-6297        & 7.0.0               & 306.87           & Clang-5-O3          & 7              & 92.80\%              & enable-zip           & 1708.22                & \xmark              & \cmark           \\
                         & CVE-2019-9025        & 7.3.0               & 378.53           & GCC-6-O0            & 5              & 99.13\%              & enable-mbstring      & 2549.78                & \xmark              & \cmark           \\
                         & CVE-2019-9638        & 7.0.0               & 305.64           & GCC-8-O3            & 6              & 85.44\%              & enable-exif          & 1729                   & \xmark              & \cmark           \\
                         & CVE-2019-9641        & 7.0.0               & 307.35           & Clang-39-O0         & 6              & 99.26\%              & enable-exif          & 1701.33                & \xmark              & \cmark           \\
                         & CVE-2019-9640        & 7.0.0               & 307.21           & GCC-7-02            & 8              & 97.01\%              & enable-exif          & 1717.82                & \xmark              & \cmark           \\
                         & CVE-2019-9639        & 7.0.0               & 309.87           & GCC-7-O1            & 8              & 98.82\%              & enable-exif          & 1712.34                & \xmark              & \cmark           \\ \hline
\multirow{5}{*}{proftpd} & CVE-2009-3639        & 1.3.1               & 9.47             & GCC-8-O1            & 7              & 99.05\%              & mod\_tls             & 17.98                  & \xmark              & \cmark           \\
                         & CVE-2010-4652        & 1.3.1               & 8.79             & Clang-6-O1          & 7              & 98.50\%              & mod\_sql             & 18.23                  & \xmark              & \cmark           \\
                         & CVE-2013-4359        & 1.3.3d              & 9.41             & GCC-6-O2            & 6              & 98.70\%              & mod\_sftp            & 31.93                  & \xmark              & \cmark           \\
                         & CVE-2015-3306        & 1.3.5b              & 10.09            & Clang-5-O3          & 7              & 98.69\%              & mod\_copy            & 39.31                  & \xmark              & \cmark           \\
                         & CVE-2016-3125        & 1.3.3d              & 9.45             & GCC-5-00            & 5              & 97.84\%              & mod\_tls             & 32.46                  & \xmark              & \cmark  \\\Xhline{1.2pt}        
\end{tabular}
\vspace{1.8pt}
\caption{Evaluation Result of Vulnerabilities in Each Stage}
\label{tab:evaluationresult}
\end{table*}

\begin{figure}[!h]
    \centering
    \includegraphics[width=8.8cm]{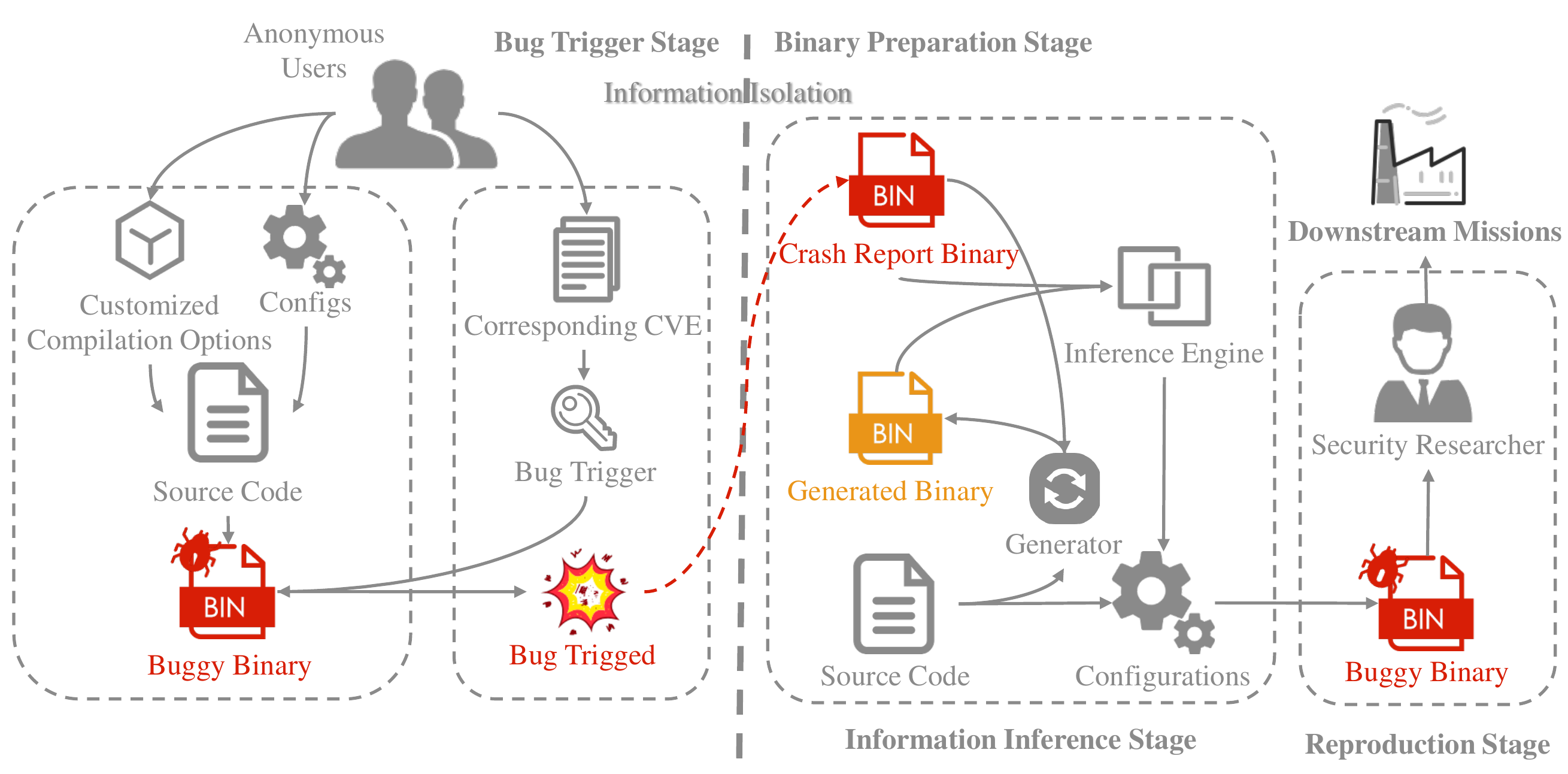}
    \vspace{-20pt}
    \caption{Procedure of Evaluation}
    \label{fig:evaluation}
\end{figure}

\textbf{Evaluation Procedures.}
As shown in Figure~\ref{fig:evaluation}, we present the procedure of setting a simulated scene for our experiment. Firstly, we divide the experiment candidates into two groups: \textit{\textbf{Anonymous Users}} (\textit{users} for short) and \textit{\textbf{Security Researchers}} (\textit{researchers} for short). Two groups of people are completely information isolated, which is consistent with what happens in the real world.

Anonymous users correspond to a variety of users in the real world. They randomly use the customized compilation options and set the program configurations according to the CVE report to compile the source code into a buggy binary. With the bug trigger, they are able to trigger the specific vulnerability. Then, they sent the bug report containing the vulnerable code segment to the researchers.

When the researchers receive the report, they are able to process the inference of compilation options and program configurations. Finally, with all the information and materials on hold, they can reproduce the failure. What's more, the result also can benefit the downstream applications of vulnerability reproduction, such as, root cause diagnosis, etc. 


\subsection{Experiment Result}

\textbf{Compilation Options Inference.}
Due to the procedure of our method, we need to recursively generate the binary. To minimize the time of building each program, we optimized the search strategy according to the analysis result of binary similarity.

In this work, we mainly focus on the vulnerabilities compiled from GCC and Clang. Firstly, \textit{users} will customize the compilation options. With the vulnerability-related program configurations, they can produce buggy binaries. After confirming the vulnerabilities can be triggered, they pack the code segment into crash reports and send them to the \textit{researchers}.

At the beginning, \textit{researchers} randomly set the default options as \textit{gcc-version6.0-O0} or \textit{clang-version5.0-O0}. They will leverage the script to automatically figure out the compilation options guided by the optimized search strategy aforementioned. The result is shown in Table~\ref{tab:evaluationresult}. Column 4 denotes the building time $t_{Build}$ in seconds, column 5 denotes the options customized by the \textit{users}, column 6 denotes the number of times to generate the binary and compare it with the crash report binary, and column 7 denotes the confidence $\beta$ of the last binary comparison produced by BinDiff. On average, we need to generate binaries about 6.6 times. In particular, with about 6 to 7 times of attempts, we can figure out the compiler, compiler version, and optimization level precisely from the binary with high confidence.

Overall, the time-consuming $t_{CO}$ of this procedure is calculated as $t_{CO} =t_{Build}\times  T_{Infer}$, which is in a wide range, from 47.23 seconds to 2478.76 seconds. The consumption of time is related to the characteristics (\eg size, complexity, etc.) of the program. And in this stage, all the customized compilation options of vulnerabilities are successfully inferred.


\textbf{Retrieval Program Configuration.}
After inferring the compilation options for each program, we retrieve the program configurations in a more fine-grained way mentioned in the previous section. According to Table~\ref{tab:evaluationresult}, column 8 denotes the vulnerability-related program configurations, and column 9 denotes the consumption of time to extract the features and map them. We spend most of our time parsing AST (Abstract Syntax Tree) of the programs. It is under our expectation that the rest of the processes take less than 20 seconds. The result also inspires us to improve the speed of parsing AST in future work.

\textbf{Configuration-Related Vulnerability Reproduction.}
With the inferred compilation options and program configurations on hold, we are able to reproduce the failures. As shown in the Table~\ref{tab:evaluationresult}, column 10 indicates that most of the vulnerabilities can not be reproduced with default options, and column 11 represents that only 2 of them can not be reproduced with the assistance of \sys due to the limited information provided. Intuitively, \sys can effectively reproduce the configuration-related vulnerabilities. Interestingly, 2 vulnerabilities from OpenSSL can be reproduced with default building configurations. But \sys still can contribute to pinpointing the vulnerable range in the default program configurations rather than other options. This ability is not available with other tools.

\subsection{Failure Cases Analysis}

According to the Table~\ref{tab:evaluationresult}, some vulnerabilities can not be reproduced by \sys. We try to figure out the reasons, which are as follows. On the one hand, the vulnerable part controlled by macro has no effect on the structure of binary code. On the other hand, BinDiff may mismatch the binary.

\vspace{-8pt}
\lstset{
    basicstyle=\scriptsize\ttfamily,
    rulesepcolor= \color{gray}, 
    breaklines=true,  
    commentstyle=\color{gray}, 
    frame=shadowbox
    }

    \begin{lstlisting}[caption={Vulnerability-related Macro in CVE-2014-3569}]
1 mask = SSL_OP_NO_TLSv1_1|SSL_OP_NO_TLSv1 
2 #if ! defined (OPENSSL_NO_SSL3) 
3           | SSL_OP_NO_SSLv3 
4 #endif 

    \end{lstlisting}
    
Take the macro above as an example. Little code is under its control so that it has almost no impact on the binary structure. When the source code is only related to the data flow but not related to the control flow, it may be directly optimized into several instructions during the compilation process. In the comparison process, it is impossible to distinguish whether the difference is caused by compiler optimization or the macro. What's more, a simple data stream may be directly optimized during the compilation process and be substituted by a result, so that there is no difference in the binary.

Secondly, if BinDiff cannot correctly find the correct relationship between the basic blocks within the two functions, it will miss the corresponding code segment controlled by the macro and lead to failure. This situation often occurs in the functions with complex jump relationships and no obvious features to distinguish the basic blocks. Meanwhile, if there is little code controlled by the macro, it will also cause a false alarm due to a slight difference. For example, the macro \textit{OPENSSL\_NO\_SSL3} only appears in 3 files in the entire OpenSSL project, and each file only has a few lines of code related to the macro.

The two situations above may cause RoBin to the failure of vulnerability reproduction, but it happens very infrequently in the real world\cite{Reproduction}. This is the limitation of our current method, and we will come out with a better solution in our future work.

%% file: related.tex
\section{Related Works} 
\label{sec:related}







\textbf{Vulnerability Reproduction.} The reproduction of vulnerabilities is quite significant to maintain the security of open-source programs. Some works~\cite{laukkanen2011survey, Reproduction} contribute to investigating the situation. One work~\cite{Reproduction} conducted an empirical analysis of real-world security vulnerabilities with the goal of quantifying their reproducibility. The presented result indicates that the vital information for reproduction is always scattered or missing due to the large gap in expert knowledge between users and developers.

\textbf{Compilation Tool-chain Inference.}
To bridge the gap between reproduction and developers, various works propose a different kind of method to identify the provenance of compilation tool-chain from the binary. Luca~\cite{massarelli2019investigating} tries to infer the provenance of the compiler by utilizing the graph neural networks to extract binary features. Bincomp~\cite{rahimian2015bincomp} proposes a hierarchical model to infer the compiler, version, and optimization levels from binary. What's more, Himalia~\cite{chen2018himalia} and its future work~\cite{yang2019understand} demonstrate the problem of distinguishing the provenance of optimization levels and make the method quite precise and efficient.

\textbf{Vulnerable Code Analysis.}
Abal~\cite{abal201442} presents a qualitative analysis of the Linux kernel for the researcher. 
TypeChef~\cite{typechef}, 
the tool also does variability-aware static analysis of software systems to detect compile and link-time errors introduced by the C preprocessor, while Vampyr~\cite{ziegler2016analyzing} proposes a static analysis of kernel drivers. And KConfigReader~\cite{kastner2017differential} analyzes the Linux kernel and transforms it into a newly presented formula.
OSSPATCHER~\cite{duan2019automating} presents a method to build the relations between OSS variants and their compiled counterparts in binaries, aiming to figure out the potential vulnerabilities. During the procedure, OSSPATCHER also tries to figure out the corresponding program configuration by the source-to-binary mapping technique.

These works mainly focus on directly analyzing source code and binary related to the vulnerabilities. In contrast, \sys gets the most of the information that causes the program to crash, trying to facilitate the reproduction of configuration-related vulnerabilities. We pay more attention to rebuild the building configurations of the vulnerabilities and try to reproduce how the users go from building open source software to cause the crash. The output materials and information are quite beneficial for the downstream applications, \eg root cause diagnosis, precise source code patch, etc.

%% file: conclusion.tex
\section{Conclusion}
\label{sec:conclusion}

In this paper, we present \sys, a comparison-based approach to figure out the building configurations to facilitate the reproduction of configuration-related vulnerabilities. We conduct an empirical study to investigate how the building configurations affect the binary on the aspect of size and similarity. With the summarized findings, \sys can infer the compilation options assisted by BinDiff. Furthermore, we overcome the challenge of inferring the program configurations by proposing a source-to-binary mapping method. Finally, the on-hold materials help to reproduce the failures and benefit the downstream applications.


%% file: journal.bbl
\begin{thebibliography}{10}

\bibitem{wannacry}
``{WannaCry Ransomware Attack}.''
  \url{https://en.wikipedia.org/wiki/WannaCry_ransomware_attack}, 2017.

\bibitem{heartbleed}
Synopsys, ``The heartbleed bug,'' 2014.

\bibitem{Reproduction}
D.~Mu, A.~Cuevas, L.~Yang, H.~Hu, X.~Xing, B.~Mao, and G.~Wang, ``Understanding
  the reproducibility of crowd-reported security vulnerabilities,'' in {\em
  27th {USENIX} Security Symposium ({USENIX} Security)}, Usenix Security '18,
  2018.

\bibitem{weidong2018osdi}
W.~Cui, X.~Ge, B.~Kasikci, B.~Niu, U.~Sharma, R.~Wang, and I.~Yun, ``{REPT}:
  Reverse debugging of failures in deployed software,'' in {\em Proceedings of
  13th {USENIX} Symposium on Operating Systems Design and Implementation
  ({OSDI})}, OSDI '18, 2018.

\bibitem{hercules}
V.-T. Pham, W.~B. Ng, K.~Rubinov, and A.~Roychoudhury, ``Hercules: Reproducing
  crashes in real-world application binaries,'' in {\em Proceedings of the 37th
  International Conference on Software Engineering (ICSE)}, ICSE ’15, 2015.

\bibitem{bugredux}
W.~Jin and A.~Orso, ``Bugredux: Reproducing field failures for in-house
  debugging,'' in {\em Proceedings of the 34th International Conference on
  Software Engineering (ICSE)}, ICSE ’12, 2012.

\bibitem{asan}
K.~Serebryany, D.~Bruening, A.~Potapenko, and D.~Vyukov, ``Addresssanitizer: A
  fast address sanity checker,'' in {\em Proceedings of the 2012 USENIX
  Conference on Annual Technical Conference}, USENIX ATC ’12, 2012.

\bibitem{debugging}
K.~Glerum, K.~Kinshumann, S.~Greenberg, G.~Aul, V.~Orgovan, G.~Nichols,
  D.~Grant, G.~Loihle, and G.~Hunt, ``Debugging in the (very) large: Ten years
  of implementation and experience,'' in {\em Proceedings of the ACM SIGOPS
  22nd Symposium on Operating Systems Principles}, SOSP ’09, 2009.

\bibitem{breakpad}
Google, ``Breakpad.''
  \url{https://github.com/google/breakpad/blob/master/docs/getting_started_with_breakpad.md},
  2007.

\bibitem{autotool}
Google, ``Gnu autotools.'' \url{https://en.wikipedia.org/wiki/GNU_Autotools},
  1992.

\bibitem{laukkanen2011survey}
E.~I. Laukkanen and M.~V. Mantyla, ``Survey reproduction of defect reporting in
  industrial software development,'' in {\em 2011 International Symposium on
  Empirical Software Engineering and Measurement}, pp.~197--206, IEEE, 2011.

\bibitem{rahimian2015bincomp}
A.~Rahimian, P.~Shirani, S.~Alrbaee, L.~Wang, and M.~Debbabi, ``Bincomp: A
  stratified approach to compiler provenance attribution,'' {\em Digital
  Investigation}, vol.~14, pp.~S146--S155, 2015.

\bibitem{massarelli2019investigating}
L.~Massarelli, G.~A. Di~Luna, F.~Petroni, L.~Querzoni, and R.~Baldoni,
  ``Investigating graph embedding neural networks with unsupervised features
  extraction for binary analysis,'' in {\em Proceedings of the 2nd Workshop on
  Binary Analysis Research (BAR)}, 2019.

\bibitem{chen2018himalia}
Y.~Chen, Z.~Shi, H.~Li, W.~Zhao, Y.~Liu, and Y.~Qiao, ``Himalia: Recovering
  compiler optimization levels from binaries by deep learning,'' in {\em
  Proceedings of SAI Intelligent Systems Conference}, pp.~35--47, Springer,
  2018.

\bibitem{yang2019understand}
S.~Yang, Z.~Shi, G.~Zhang, M.~Li, Y.~Ma, and L.~Sun, ``Understand code style:
  Efficient cnn-based compiler optimization recognition system,'' in {\em ICC
  2019-2019 IEEE International Conference on Communications (ICC)}, pp.~1--6,
  IEEE, 2019.

\bibitem{rosenblum2011recovering}
N.~Rosenblum, B.~P. Miller, and X.~Zhu, ``Recovering the toolchain provenance
  of binary code,'' in {\em Proceedings of the 2011 International Symposium on
  Software Testing and Analysis}, pp.~100--110, 2011.

\bibitem{duandeepbindiff}
Y.~Duan, X.~Li, J.~Wang, and H.~Yin, ``Deepbindiff: Learning program-wide code
  representations for binary diffing,'' 2020.

\bibitem{liu2018alphadiff}
B.~Liu, W.~Huo, C.~Zhang, W.~Li, F.~Li, A.~Piao, and W.~Zou, ``$\alpha$diff:
  cross-version binary code similarity detection with dnn,'' in {\em
  Proceedings of the 33rd ACM/IEEE International Conference on Automated
  Software Engineering}, pp.~667--678, 2018.

\bibitem{xu2017neuralSIM}
X.~Xu, C.~Liu, Q.~Feng, H.~Yin, L.~Song, and D.~Song, ``Neural network-based
  graph embedding for cross-platform binary code similarity detection,'' in
  {\em Proceedings of the 2017 ACM SIGSAC Conference on Computer and
  Communications Security}, pp.~363--376, 2017.

\bibitem{zuo2018neuralINNEREYE}
F.~Zuo, X.~Li, P.~Young, L.~Luo, Q.~Zeng, and Z.~Zhang, ``Neural machine
  translation inspired binary code similarity comparison beyond function
  pairs,'' {\em arXiv preprint arXiv:1808.04706}, 2018.

\bibitem{bindiff}
Google, ``Bindiff.'' \url{https://www.zynamics.com/bindiff.html}, 2016.

\bibitem{Flake2004StructuralCO}
H.~Flake, ``Structural comparison of executable objects,'' in {\em Proceedings
  of the IEEE Conference on Detection of Intrusions and Malware \&
  Vulnerability Assessment (DIMVA)}, DIMVA '04, 2004.

\bibitem{Dullien2005GraphbasedCO}
T.~Dullien and R.~Rolles, ``Graph-based comparison of executable objects,'' in
  {\em Symposium sur la securite des Technologies de l'information et des
  Communications (SSTIC)}, SSTIC '05, 2005.

\bibitem{disassembler2010debugger}
Hex-Rays, ``Disassembler, ida pro,'' 2010.

\bibitem{dong2019towards}
Y.~Dong, W.~Guo, Y.~Chen, X.~Xing, Y.~Zhang, and G.~Wang, ``Towards the
  detection of inconsistencies in public security vulnerability reports,'' in
  {\em 28th $\{$USENIX$\}$ Security Symposium ($\{$USENIX$\}$ Security 19)},
  pp.~869--885, 2019.

\bibitem{abal201442}
I.~Abal, C.~Brabrand, and A.~Wasowski, ``42 variability bugs in the linux
  kernel: a qualitative analysis,'' in {\em Proceedings of the 29th ACM/IEEE
  international conference on Automated software engineering (ASE)}, ASE ’14,
  2014.

\bibitem{typechef}
A.~Kenner, C.~K\"{a}stner, S.~Haase, and T.~Leich, ``Typechef: Toward type
  checking \#ifdef variability in c,'' in {\em Proceedings of the 2nd
  International Workshop on Feature-Oriented Software Development (FOSD)}, FOSD
  '10, 2010.

\bibitem{ziegler2016analyzing}
A.~Ziegler, V.~Rothberg, and D.~Lohmann, ``Analyzing the impact of feature
  changes in linux,'' in {\em Proceedings of the Tenth International Workshop
  on Variability Modelling of Software-intensive Systems}, pp.~25--32, 2016.

\bibitem{kastner2017differential}
C.~K{\"a}stner, ``Differential testing for variational analyses: Experience
  from developing kconfigreader,'' {\em arXiv preprint arXiv:1706.09357}, 2017.

\bibitem{duan2019automating}
R.~Duan, A.~Bijlani, Y.~Ji, O.~Alrawi, Y.~Xiong, M.~Ike, B.~Saltaformaggio, and
  W.~Lee, ``Automating patching of vulnerable open-source software versions in
  application binaries.,'' in {\em NDSS}, 2019.

\end{thebibliography}
